\documentclass[%
 reprint,
 superscriptaddress,
 amsmath,amssymb,
 aps,
 prl, 
]{revtex4-2}
\usepackage{physics}
\usepackage{graphicx}
\usepackage{dcolumn}
\usepackage{bm}
\usepackage[colorlinks=true,
            linkcolor=red,
            citecolor=blue,
            urlcolor=blue,
            pdfborder={0 0 0}]{hyperref}

\begin{document}

\title{Optimized spectral and interferometric techniques for the certification of ETPA}

\author{Pablo Yepiz-Graciano}
\email{pablodyg@gmail.com}
\affiliation{Instituto de Ciencias Nucleares, Universidad Nacional Autónoma de México, CDMX, 04510, México.}

\author{Gabriel Ramos-Ortiz}
\email{garamoso@cio.mx}
\affiliation{Centro de Investigaciones en Óptica, A.C., Loma del Bosque 115, Colonia Lomas del Campestre, 37150 León, Guanajuato, México.}

\author{Roberto Ramírez-Alarcón}
\email{roberto.ramirez@cio.mx}
\affiliation{Centro de Investigaciones en Óptica, A.C., Loma del Bosque 115, Colonia Lomas del Campestre, 37150 León, Guanajuato, México.}


\begin{abstract}
The phenomenon of Entangled Two-Photon Absorption (ETPA) presents a persistent controversy in the literature, evidenced by a wide disparity in the reported values for the $\sigma_E$ cross-sections. Much of this discrepancy is attributed to the difficulty in discriminating ETPA from various background processes that can mimic its signal, such as linear absorption or scattering. Given this need to certify the presence of ETPA unequivocally, this work introduces a key strategy to isolate the ETPA contribution through its spectral signature. This involves modeling the molecule as a two-photon notch filter and inducing a controlled asymmetric overlap with the joint spectral intensity (JSI) of the incident photons. This asymmetry is used to generate a measurable distortion in the transmitted JSI, and, complementarily, as a reduction in the visibility of the Hong-Ou-Mandel (HOM) dip. To ensure the experimental feasibility of this technique, a comprehensive analysis of the experimental conditions required for its detection is presented, establishing the absorption efficiency limits that must be overcome given the constraints of entangled-photon-pair flux and detector noise. A preliminary experiment addressing the ETPA detection limits using the standard RhB dye is presented.
\end{abstract}

\maketitle

\section{Introducción}
The process of two-photon absorption (TPA), first described theoretically in 1931 by Maria Goppert Mayer, involves the simultaneous absorption of two photons that occurs when the sum of their energies ($E_1+E_2$) is in resonance with one of the electronic states of a molecule. As a result of the absorption process, the two photons are lost from the excitation beam, reducing its intensity, while the molecule is brought to an excited state that later decays to the ground state by either a radiative or non-radiative process \cite{MariacristinaRumi2010}. This TPA effect finds many applications, to name a few, three-dimensional optical storage memory \cite{Dimitri1989,Strickler1991}, axial super-resolution for microfabrication \cite{Wei2007} and ultra-high resolution microscopy \cite{WDenk1990,Pastirk2003,HELL1992}.
To be observed, the TPA effect requires high intensities of excitation, which are obtained with laser beams of short pulses which are tightly focused into a nonlinear sample; these conditions of high laser excitation are disadvantageous in the study of biological systems and other delicate systems. \cite{PATTERSON2000}. In recent times, however, a very interesting possibility has arisen to generate TPA using entangled photon pairs (EPP) produced through spontaneous parametric down-conversion (SPDC)  \cite{Javanainen1990}. This non-conventional approach opens up the possibility of using much lower photon fluxes since the ETPA cross-sections  ($\sigma_E$) can be orders of magnitude greater than the TPA cross-sections ($\sigma_C$).

The increasing interest in ETPA during the last decade has conducted to multiple studies on organic chromophores \cite{Upton2013}, dendrimers \cite{Guzman2010}, thianocenes \cite{Eshun2018}, Zn-TPP porphyrin \cite{Varnavski2023Colorsentangled}, and laser dyes mainly Rhodamine B (RhB) \cite{Villabona2017} and Rhodamine 6G (Rh6G) \cite{He2024,Parzuchowski2021}, along with theoretical investigations of the electronic structure of chromophores\cite{Kang2020, Oka2020}. Various techniques have been proposed to control the process, such as the Spatial Control of ETPA \cite{Guzman2010}, picoseconds spectroscopy \cite{Burdick2021}, and variable pump spectroscopy \cite{Mertenskotter2021}. Additionally, studies on the effect of multi-chromatic superposition and its influence on molecular transitions have been carried out \cite{Wittkop2023}, among other applications \cite{Smith2024, Schlawin2022, Gu2021}.

In the recent investigations on the ETPA phenomena some inconsistencies have arisen between theoretical predictions and experimental results, and even among different experimental techniques, raising questions such as: What is the real order of magnitude of the  $\sigma_E$ cross section? Reported values range from $10^{-11}$ cm$^2$/molecule in atomic systems \cite{Fei1997} to values in molecular systems that vary from $10^{-17}$ to $10^{-19}$ cm$^2$/molecule \cite{Upton2013, Guzman2010, Eshun2018, Kang2020, Burdick2021, Villabona2017}. However, more detailed analysis suggests that reported experimental values may be overestimated because effects that mimic the ETPA are not properly discriminated. Some effects that mimic ETPA are scattering, linear absorption, and hot-band absorption\cite{Mikhaylov2022}. Therefore, some studies suggest that the actual $\sigma_E$ values should be orders of magnitude lower, in the range $10^{-22}$ - $10^{-25}$ cm$^2$/molecule \cite{Tabakaev2021, Parzuchowski2021, He2024}, thus questioning the feasibility of quantum enhancement of the nonlinear absorption\cite{Raymer2021, Landes2021}. Different works have also discussed the limits of detection of the ETPA process considering the attainable levels of EPP fluxes and available detectors \cite{Parzuchowski2021, Landes2021_ExperimentalFeasibility, Landes2024, He2024}.

Given that multiple processes can overlap or combine to mimic ETPA with similar strength — or even obscure it entirely — researchers have sought unmistakable characteristics unique to the ETPA process. It was originally thought that the linear growth of the ETPA rate with EPP flux was a sufficient and distinctive feature for studying this effect. However, as mentioned earlier, linear processes that mimic ETPA can follow the same pattern \cite{Dayan2005,Mikhaylov2022,Landes2024}. Other techniques suggest using spatial correlation with a Z-scan and observing the effect as it exits the focus zone \cite{Tabakaev2022SpatialProperties}.
The correlation time of the bi-photon has been also proposed for turning on/off the ETPA process when the twin photons are overlapped/separated beyond their coherence time \cite{Tabakaev2021, Corona-Aquino2022}.
Spectral techniques comprising a tunable pump laser are used to measure the cross sections for different energy ranges of the EPP, allowing differentiation between signals of one-photon resonant absorption (OPA), ETPA, and classical TPA \cite{Varnavski2023Colorsentangled}.
Interferometric techniques, such as placing the sample in one arm of a Hong-Ou-Mandel (HOM) interferometer in order to examine the changes in the interference, have been tested \cite{Eshun2021HOM}. Further, it was demonstrated that placing the sample in a collinear HOM interferometer produces HOM interferograms insensitive to linear losses \cite{Triana-Arango2024}.
Even theoretical investigations have proposed the use of NOON state interferometers, which are also insensitive to linear losses \cite{Martinez-Tapia2023}.

In a recent work from our group (Triana-Freiman \textit{et al.} \cite{Triana-Arango2023}), the effect of ETPA on the the visibility of the HOM dip was studied from spectral considerations, where the spectral profile of the molecular absorption was modeled as a notch filter acting on the joint spectral intensity (JSI) function. However, the proposed conditions to observe a change in visibility required the combination of a pump of wide spectral bandwidth with a hypothetical molecule having a very narrow  and efficient two-photon absorption band —conditions hard to meet for real molecules. In a related work by  Martínez-Tapia \textit{et al.} \cite{Martinez-Tapia2023} the notch filter model was also used to compare how different configurations for the two-photon state behave after the ETPA interaction, concluding that a type-II SPDC state or a two-photons NOON state were the best candidates to certify the occurrence of ETPA. In this sense, it is important to mention that by modeling the molecule as a notch filter, which fulfills frequency conservation, it is possible to observe the occurrence of the ETPA process as a direct effect over the joint spectral intensity (JSI) function of the SPDC two-photon state. This change of perspective permits us to consider novel sensing schemes capable of certifying ETPA, for instance direct measurements of the JSI or changes in the visibility or the width of the Hong-Ou-Mandel dip, produced by the photon pairs after interacting with the molecule under study.

In this direction, in the present work we focus on developing a strategy for certificate the occurrence of ETPA under realistic experimental conditions, based on identifying the spectral signature that the absorption process produce over the JSI function of the EPP. To produce a measurable change over the JSI function, the JSI must be asymmetrically overlapped in the frequency space with the molecular absorption band, modeled as a two-photon notch filter. The key to achieve this asymmetrical overlap is to produce photon pairs with an asymmetric JSI function, which is possible exploiting the strong birefringence of a periodically-poled potassium titanyl phosphate (PPKTP) crystal when pumped with a broad spectra femtosecond pulsed laser. To measure the ETPA interaction we propose two techniques: (1) Spectral detection by directly measuring the JSI function and observing the modification of the biphoton's spectral correlations. (2) Interferometric detection with the HOM interferometer, which due to the loss of spectral indistinguishability, caused by the asymmetric absorption, results in a measurable reduction in the visibility and changes in the width of the HOM dip. Subsequently, the detection limit of the technique is established through a quantitative evaluation of the experimental conditions, which includes simulating the number of EPPs generated within the JSI and considering detector noise limitations, which allows for the determination of the necessary threshold absorption efficiency for the ETPA process to be observable. This robust approach provides a complementary pathway for the certification of ETPA occurrence.
Finally, a preliminary experiment addressing the ETPA detection limits using the standard RhB dye is presented by calculating the pair flux level generated in a ppKTP crystal pumped by a pulsed laser and considering the current noise level of conventional silicon avalanche single-photon detectors.

\section{\label{sec:Theory}Theory}

\subsection{\label{subsec:Spectral and Interferometric Technique}Spectral and Interferometric Technique for ETPA certification}

Consider the EPP generated through SPDC, which is described by the state $\ket{\psi}=\ket{0}_1 \ket{0}_2+\xi \int d\omega_s \int d\omega_i f(\omega_s,\omega_i) \ket{\omega_s} \ket{\omega_i}$, where $\xi$ is a constant related to the conversion efficiency, $f(\omega_s,\omega_i)$ is the joint spectral amplitude (JSA) function, and its squared modulus $\abs{f(\omega_s,\omega_i)}^2$ is the so-called joint spectral intensity (JSI) function, which is equivalent to the probability of producing a s-photon with frequency $\omega_s$ and an i-photon with frequency $\omega_i$.

The JSA is equal to
\begin{equation}\label{eq:jsi}
    f(\omega_s,\omega_i)=\alpha(\omega_s,\omega_i)\Phi(\omega_s,\omega_i),
\end{equation}

\noindent where $\Phi(\omega_s,\omega_i)=\text{sinc} \left(\frac{L}{2} \Delta k\right)$ is the phase matching function, $L$ the crystal length and $\Delta k =
k_{p} -k_{sz}-k_{iz}-\frac{2\pi}{\Lambda}$ is the collinear phase mismatch, that is, how much the system deviates from exactly satisfying the momentum conservation; $k_{p}$, $k_{sz}$, $k_{iz}$ are the longitudinal components of the wavevectors of the pump, signal and idler photons, respectively; $\Lambda$ is the period of a periodically poled crystal (ppKTP) and $\alpha(\omega_{p})=\exp{\left[- \frac{(\omega_{p}-\omega_{p0})^{2}}{2\sigma_{p}^{2}}\right]}$ is the pump spectral envelope, where $\omega_{p0}$ and $\sigma_{p}$ are the central frequency and bandwidth of the pump, respectively. Evidently, by conservation of energy $\omega_{p}$=$\omega_s$+$\omega_i$.

\begin{figure*}[t]
	\centering
	\includegraphics[width=.8\textwidth]{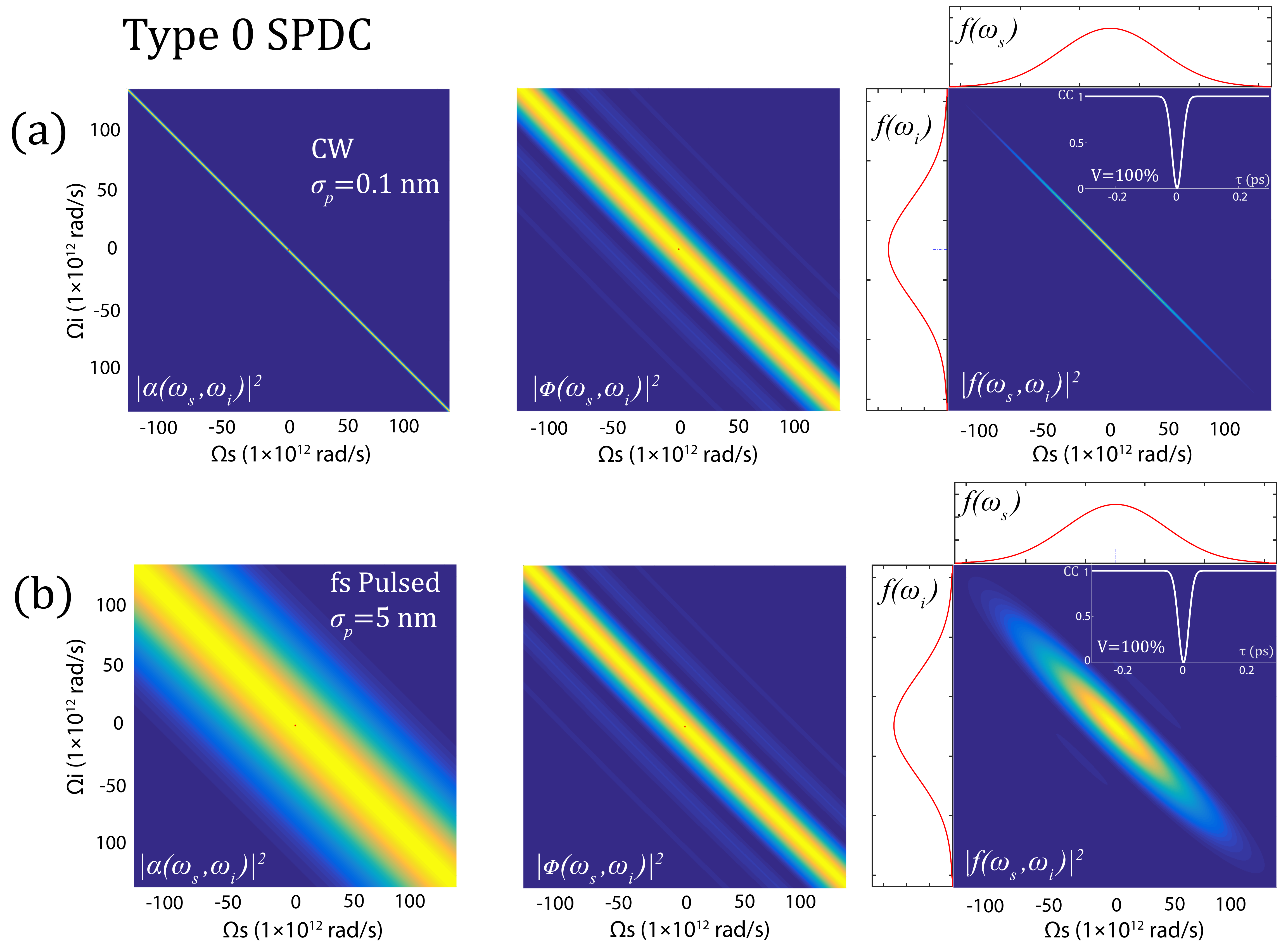}
	\caption{$\abs{\alpha}^2$ function (first column) and the $\abs{\Phi}^2$ function (second column), with the third column showing the JSI $\abs{f(\omega_s,\omega_i)}^2$ and its marginals $f(\omega_s)$, $f(\omega_i)$ for type-0 SPDC generated in a ppKTP crytal (L=10 mm). An inset was added showing the HOM dip and the visibility of the resulting interferogram in the cases of employing: (a) a continuous wave laser ($\sigma_p=0.1$ nm); (b) a pulsed laser ($\sigma_p=5$ nm). [The axes were rescaled using $\Omega_{\mu}=\omega_{\mu}-\omega_{0}$, with $\omega_0=\omega_{p0}/2$ and $\mu$ = $s$ or $i$]} 
	\label{fig: alpha_phi_jsi_comp_type0}
\end{figure*}

Thus, the JSI is limited by the product of the spectral envelope, $\alpha$, and the longitudinal phase-matching function, $\Phi$, as shown in Figure \ref{fig: alpha_phi_jsi_comp_type0} for the case of Type 0 SPDC in a ppKTP crystal of $L=10$ mm, and period $\Lambda=10$ $\mu$m; the figure considers the cases of continuous-wave (CW) and pulsed pump, with a central wavelength of $\lambda_{p0}=405$ nm ($\omega_{p0}=4.65\times10^{15}$ rad/s), with characteristic bandwidths $\sigma_p$ of 0.1 nm and 5 nm, respectively. This figure also presents the marginals $f(\omega_s)=\int d\omega_i\abs{f(\omega_s,\omega_i)}^2 $ and  $f(\omega_i)=\int d\omega_s \abs{f(\omega_s,\omega_i)}^2$.  

The coincidence rate $R_C(\tau)$ of the Hong-Ou-Mandel interferogram as a function of the time delay between the signal and idler photons when they reach the beam splitter in a HOM interferometer, directly depends on the JSI in the following way:
\begin{equation}\label{eq:hom}
    R_C(\tau)=\frac{1}{4}\int \int d\omega_s d\omega_i \abs{f(\omega_s,\omega_i)e^{i(\omega_s-\omega_i)\tau}-f(\omega_i,\omega_s)}^2,
\end{equation}

\noindent where $\Omega = \omega_s - \omega_i$ is the detuning of the photons. The maximum value this variable can reach is $\Omega^{Max}=\sigma_a$, where $\sigma_a$ is the anti-diagonal width of the JSI which is inversely proportional to the width of the HOM dip. The visibility of the HOM dip is calculated as $V=(R_C(\infty)-R_C(0))/(R_C(\infty)+R_C(0))$, where $R_C(\infty)$ are the coincidences outside the dip, and $R_C(0)$ are the coincidences within the dip.

When considering $\tau = 0$, the minimum coincidence rate in the HOM interferogram is reached, which will be $R_C(0)=0$ (or equivalently $V=100$\%) only if the state is spectrally symmetric, namely $f(\omega_s, \omega_i) = f(\omega_i, \omega_s)$. This can be visually confirmed by exchanging the frequency axes, where the distribution remains unchanged. Figure \ref{fig: alpha_phi_jsi_comp_type0} shows how, when using a SPDC process of type-0, in which both photons have the same polarization, their spectral components undergo the same changes, and their spectral distribution is ideally similar and symmetric.

\begin{figure*}[t]
	\centering
	\includegraphics[width=0.8\textwidth]{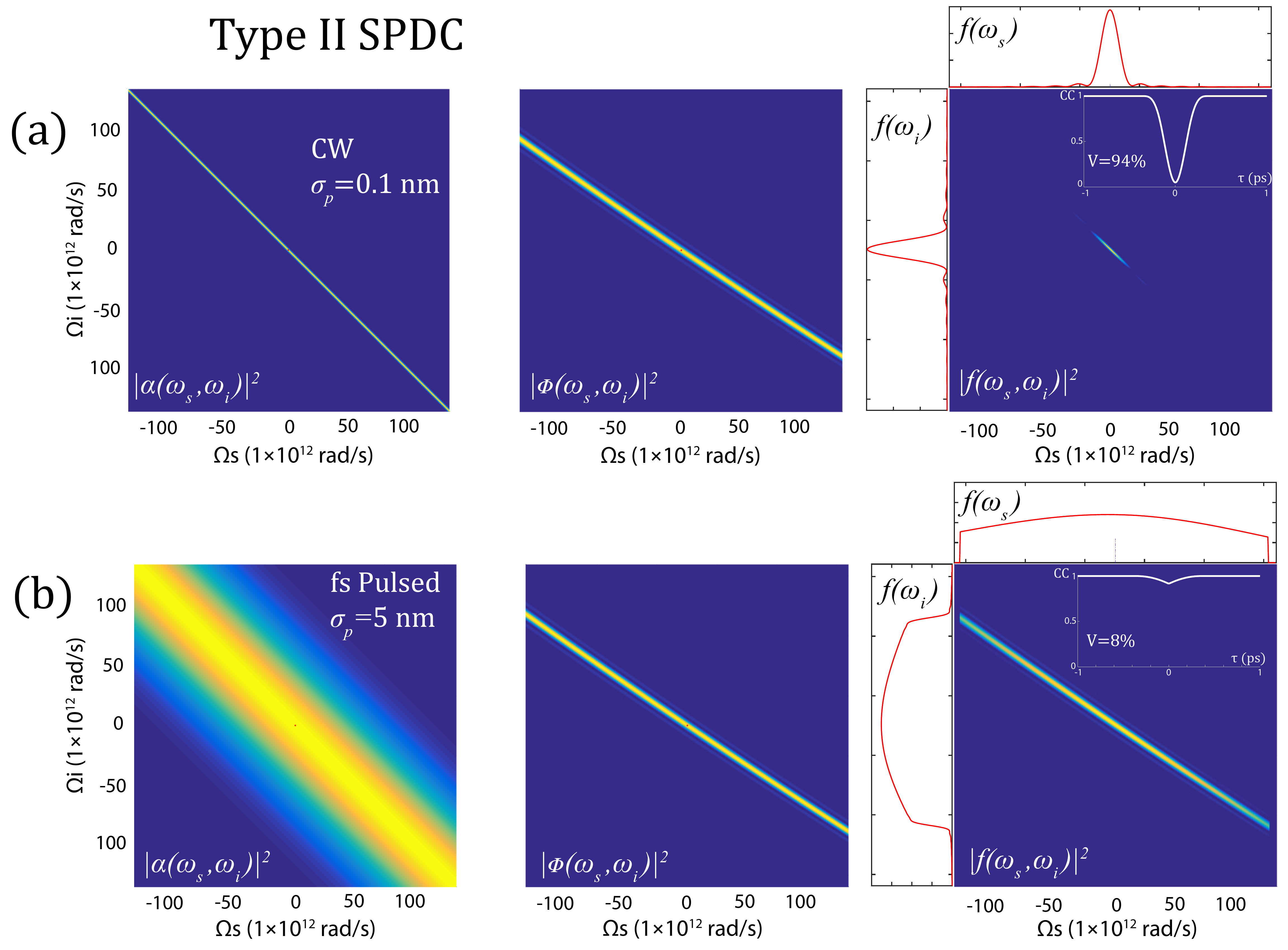}
	\caption{JSI function for type-II SPDC generated in a ppKTP crytal (L=10 mm) as the product of the $\abs{\alpha}^2$ function (first column) and the $\abs{\Phi}^2$ function (second column). The third column shows the JSI $\abs{f(\omega_s, \omega_i)}^2$ and its marginals $f(\omega_s)$ and $f(\omega_i)$. An inset was added showing the HOM dip and visibility. (a) In the case of using a CW laser ($\sigma_p = 0.1$ nm), the HOM dip reaches $V=94\%$ because, although the $\Phi$ function is highly tilted, the narrow diagonal width of the pump is still cut symmetrically such that the marginal distributions are quite similar. (b) In the case of using a pulsed laser ($\sigma_p = 5$ nm), the JSI loses its symmetry, and it is immediately clear that the marginal distributions are very different. It is observed that visibility is almost completely lost, $V = 8\%$. [The axes were rescaled using $\Omega_{\mu} = \omega_{\mu} - \omega_0$, with $\omega_0 = \omega_{p0}/ 2$ and $\mu$ = $s$ or $i$].} 
	\label{fig: alpha_phi_jsi_comp_type2}
\end{figure*} 

Figure \ref{fig: alpha_phi_jsi_comp_type2} shows the case of a type-II SPDC process, considering again $\lambda_{p0}=405$ nm ($\omega_{p0}=4.65\times10^{15}$ rad/s), in which the signal and idler photons have orthogonal polarizations, causing the phase-matching function $\Phi$ to be no longer symmetric, with a tilt relative to the perfect anti-diagonal, making the JSI asymmetric such that in general the marginals follow the relation $f(\omega_s) \neq f(\omega_i)$. If we use a CW laser, we can artificially symmetrize the marginals, $f(\omega_s) \approx f(\omega_i)$, as shown in the third column of Fig. \ref{fig: alpha_phi_jsi_comp_type2}(a) with a small decrease of visibility in a HOM interferometer reaching $V$=$94$\%. However, if we use a pump pulse of broadband spectrum, the asymmetry in the JSI becomes noticeable, and the direct effect of visibility loss in the HOM interferogram is evident, with $V$=$8\%$, as shown in Fig. \ref{fig: alpha_phi_jsi_comp_type2}(b).

We model the effect of the two-photon absorption process on the JSI function, assuming that the spectral profile of sample's absorption behaves as a Notch filter, $h_{N}(\omega_s,\omega_s)=1-\eta \exp{\left[- \frac{(\omega_{s}+\omega_{i}-2\Omega_{N_0})^{2}}{2\sigma_{N}^{2}}\right]}$, where $\eta$ is the absorption efficiency, while $\Omega_{N_0}$ and $\sigma_N$ are the central absorption frequency and bandwidth of the filter, respectively.
The absorption efficiency is simply $\eta=\frac{N{abs}}{N_{pairs}}$, where $N_{pairs}$ is the total number of photon pairs passing through the sample, and $N_{abs}$ is the number of pairs absorbed by the sample. Logically, the value is constrained between $0 \leq \eta \leq 1$, where a value of zero means that no pairs were absorbed ($N_{abs}=0$), and one means that all incident pairs were absorbed ($N_{abs}=N_{pairs}$).

As the photons pass through the sample and considering that some of these photons will be absorbed, we can write the JSA as follows:
\begin{eqnarray}    
f_{sam}(\omega_s,\omega_i)&=&
f(\omega_s,\omega_i)h_{N}(\omega_s,\omega_i) \nonumber \\ 
&=&\alpha(\omega_s,\omega_i)\Phi(\omega_s,\omega_i)h_{N}(\omega_s,\omega_i).
\end{eqnarray}

\begin{figure*}[t]
	\centering
	\includegraphics[width=0.8\textwidth]{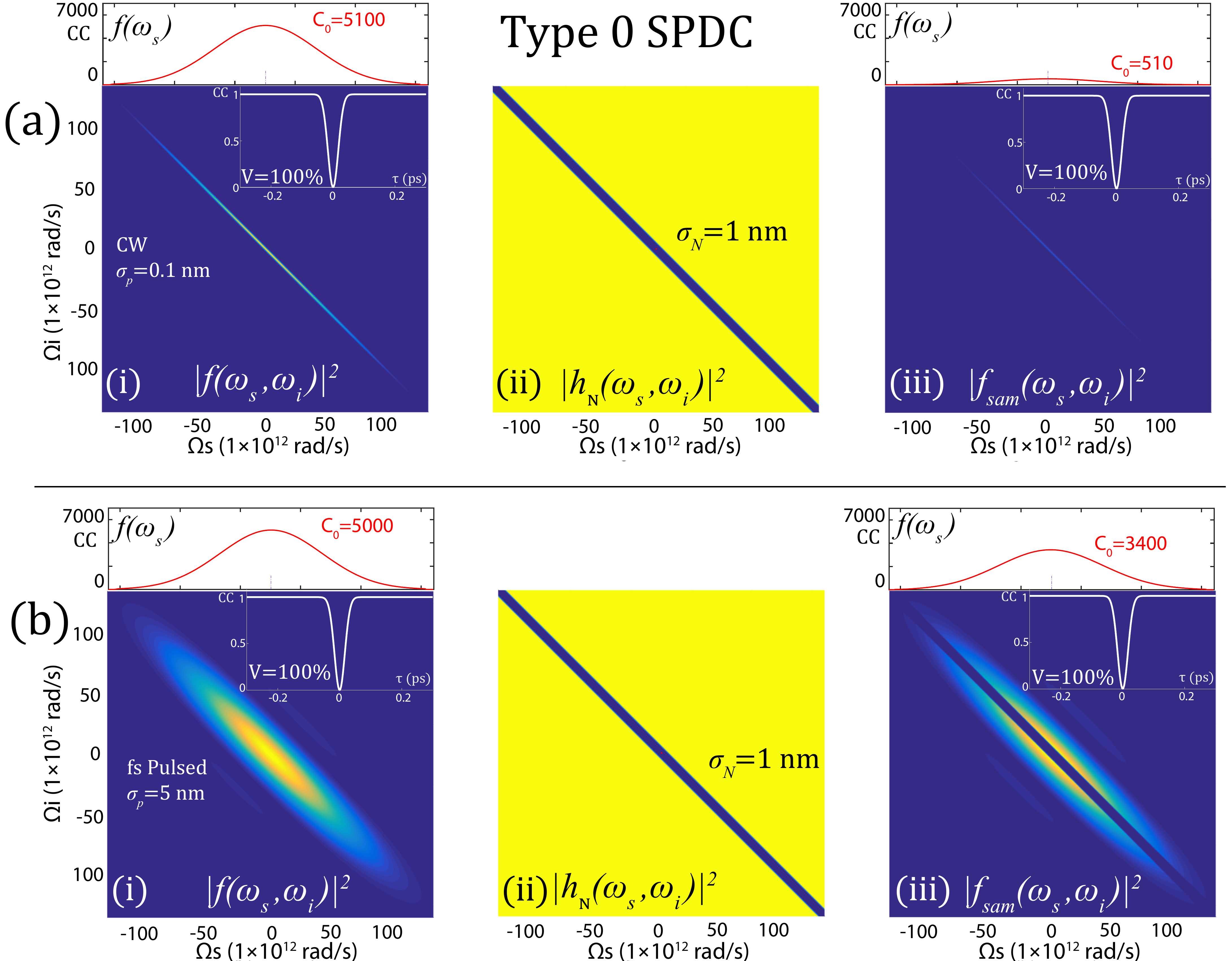}
	\caption{Effect  generated for a type-0 SPDC process on the JSI, $\abs{f(\omega_s,\omega_i)}^2$, and on the HOM visibility interferogram, $R_C(\tau)$, before and after being filtered by a Notch filter. The top row (a) refers to the narrow bandwidth CW case and the bottom row (b) refers to the broad bandwidth case (femtosecond pulsed laser). (i) JSI, $\abs{f(\omega_s,\omega_i)}^2$, marginal $f(\omega_s)$ (red), and the interferogram (white) along with the corresponding visibility. (ii) The Notch filter, $\abs{h_N(\omega_s,\omega_i)}^2$, which forms an anti-diagonal with width, $\sigma_N=1$ nm, and efficiency, $\eta=0.90$. (iii) The JSI modified by the Notch filter, $\abs{f_{sam}}^2=\abs{f}^2 \times \abs{h_N}^2$
[The axes were rescaled using $\Omega_{\mu}=\omega_{\mu}-\omega_{0}$, where $\omega_0=\omega_{p0}/2$]
[The vertical marginal was omitted since it is symmetric for this process, $f(\omega_i)=f(\omega_s)$.]}
	\label{fig: jsi_notch_comp_type0}
\end{figure*}

\begin{figure*}[t]
	\centering
	\includegraphics[width=0.8\textwidth]{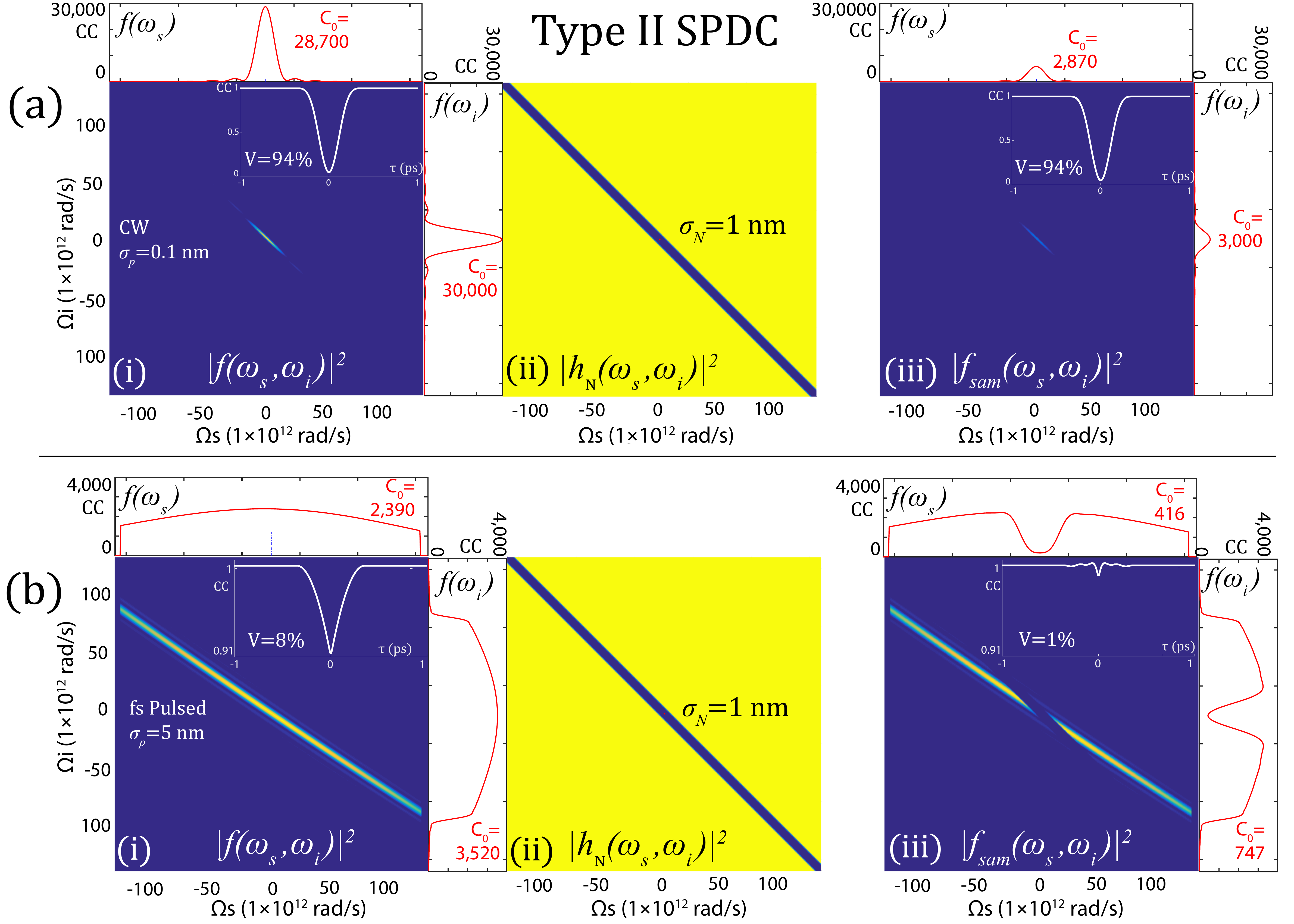}
	\caption{Effect for a type-II SPDC process on the JSI, $\abs{f(\omega_s,\omega_i)}^2$. The insets show the effect on $R_C(\tau)$, before and after being filtered by a Notch filter. The top column (a) refers to the CW case and the bottom column (b) refers to the pulsed laser case. (i) JSI, $\abs{f(\omega_s,\omega_i)}^2$, marginal $f(\omega_s)$ (red), and the interferogram (white) along with the corresponding visibility. (ii) The Notch filter, $\abs{h_N(\omega_s,\omega_i)}^2$, which forms an anti-diagonal, $\sigma_N=1$ nm, and efficiency, $\eta=0.90$. (iii) JSI modified by the Notch filter, $\abs{f_{sam}}^2=\abs{f}^2 \times \abs{h_N}^2$
[The axes were rescaled using $\Omega_{\mu}=\omega_{\mu}-\omega_{0}$, where $\omega_0=\omega_{p}/2$]
[The vertical marginal was omitted since it is symmetric for this process, $f(\omega_i)=f(\omega_s)$.]}
	\label{fig: jsi_notch_comp_type2}
\end{figure*}

We illustrate the effect of an idealized Notch filter on the HOM interferogram in Figure \ref{fig: jsi_notch_comp_type0} and Figure \ref{fig: jsi_notch_comp_type2}, corresponding to type-0 and a type-II SPDC process, respectively, under the condition $\lambda_{N0}=810$ nm ($\Omega_{N_0}= 2.32\times10^{15}$ rad/s). Here the idealized Notch filter has a narrow bandwidth of $\sigma_N = 1$ nm and efficiency of $\eta$ = 0.9. Naturally, such a high level of nonlinear absorption implies that the ETPA could be detected easily and unequivocally in simple transmission experiments. However, here the parameter values were chosen arbitrarily, solely for illustrative purposes to demonstrate the proposed method for certifyng ETPA.  

In the case of a symmetric JSI function, generated through a SPDC type-0 process (Figure \ref{fig: jsi_notch_comp_type0}), there are two extreme cases: (a) the case of a narrow pump bandwidth (CW) and (b) a broad pump bandwidth (pulsed). For the CW case, with a pump bandwidth of $\sigma_p=0.1$ nm, the effect of an idealized Notch filter with an arbitrary narrow bandwidth of $\sigma_N=1$ nm is a reduction of the entire distribution, as shown by the 90\% attenuation of coincidence counts in the marginals. For instance, for $f(\omega_s)$ decreases from $5,100$ CC to $510$ CC. Since the process is symmetric, it is understood that both marginals are equal ($f(\omega_s)=f(\omega_i)$). As the symmetry does not change, since the loss is the same for both axes, the final JSI function remains symmetric ($f(\omega_s,\omega_i)=f(\omega_i,\omega_s)$). Therefore, the HOM interferogram remains unchanged.

For the pulsed case, $\sigma_p=5$ nm (Figure \ref{fig: jsi_notch_comp_type0}(b)), the JSI function is much more spread along the diagonal, and the Notch filter cuts through it, creating a hole along the diagonal. The effect on the marginals is again symmetric, meaning that the losses on each axis are the same, so at the end $f(\omega_s)=f(\omega_i)$. The values of the marginal decrease from approximately 5 000 CC to 3 400 CC. Again the absorption effect is null on the visibility of the HOM interferogram despite the ETPA process was $90$\% efficient. Therefore, a type-0 or type-I SPDC process is ruled out as an option for detecting ETPA through the HOM interferometric technique.

In the case of an asymmetric JSI, generated through a type-II process, (Figure \ref{fig: jsi_notch_comp_type2}), we again present the extreme cases of narrow bandwidth (CW) and broad bandwidth (pulsed) pumping. For the CW case, as already mentioned and shown in Figure \ref{fig: alpha_phi_jsi_comp_type2}(a), the effect of a narrow-band pump laser leads to a nearly symmetric JSI; the weak asymmetry of the JSI here traduces only in a moderate loss of visibility, $V$=94\%, where $f(\omega_s)\approx f(\omega_i)$. If we now apply a 90\%-efficiency Notch filter with narrow bandwidth of 1 nm to this state \ref{fig: alpha_phi_jsi_comp_type2}(aiii), we observe that the JSI at the output is only attenuated, with the marginal $f(\omega_s)$ counts dropping from $28,700$ C to $2,870$ C, and with the marginal $f(\omega_i)$ counts dropping from $30,000$ C to $3,000$ C, preserving the pre-existing symmetry and thus not modifying the HOM visibility.

Finally, for the pulsed case, $\sigma_p=5$ nm (Figure \ref{fig: jsi_notch_comp_type2}(b)), the JSI is slightly broader along the diagonal and significantly more extended along the anti-diagonal, and we can see how the Notch filter cuts through it, creating a hole along the main anti-diagonal. This time, the cut is not symmetric; although the Notch filter cuts symmetrically, the JSI is tilted (approximately $11^\circ$). The effect of the cut on the marginals is very noticeable, which means that the losses on each axis are distinct, so $f(\omega_s) \neq f(\omega_i)$, and it produces a radical change in the HOM dip visibility from 8\% to 1\%. Therefore, the type-II SPDC process with pulsed pump is ideal benchmark for detecting the ETPA process through the HOM interferometric technique.

\subsection{Notch filter bandwidth and frequency detuning effects on ETPA detection}

\begin{figure*}[t]
	\centering
	\includegraphics[width=0.8\textwidth]{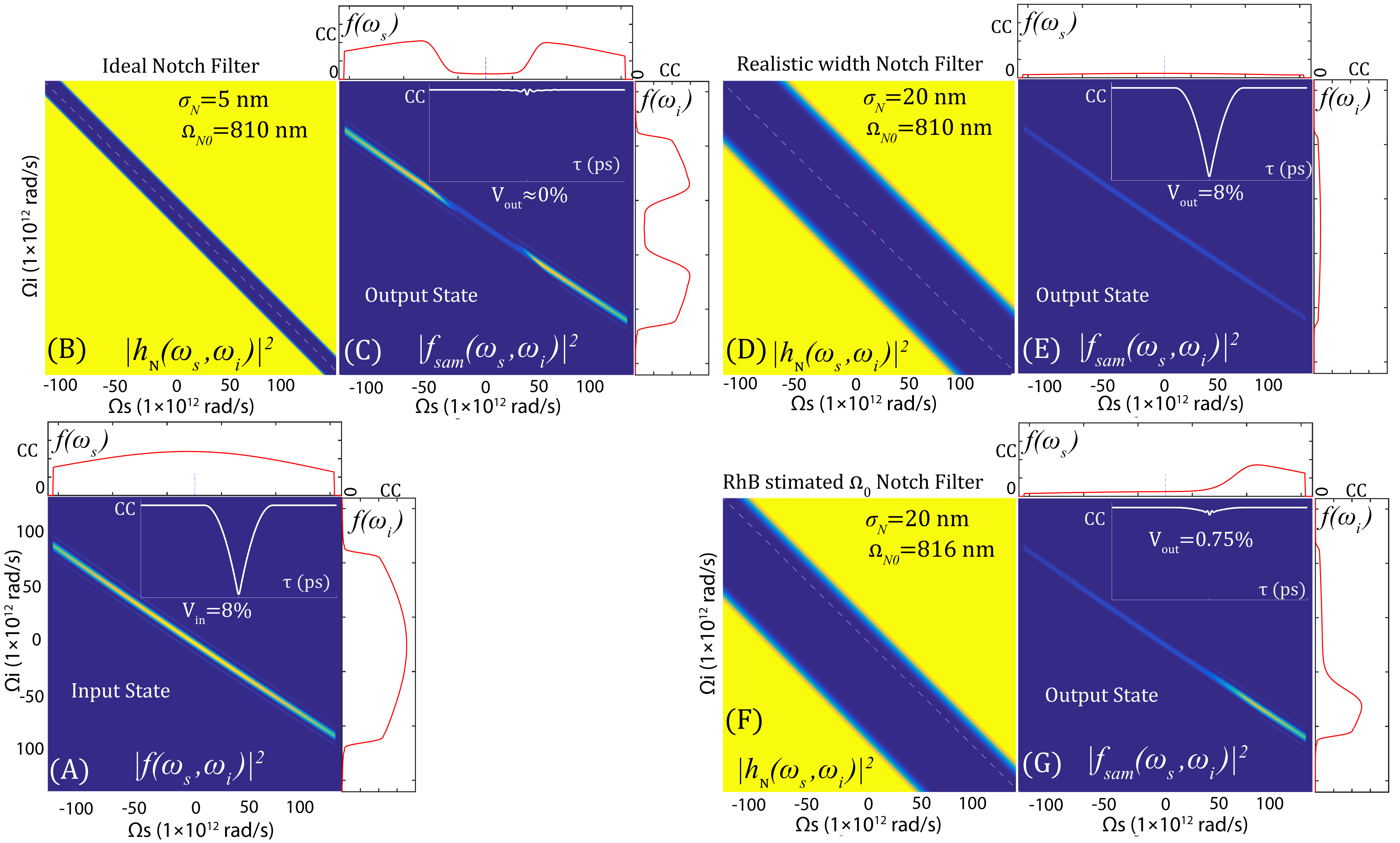}
	\caption{Realistic bandwidths for Notch filters. We selected the best JSI case, generated by a type-II SPDC process with a pump bandwidth of $\sigma_p = 5$ nm, corresponding to Figure \ref{fig: jsi_notch_comp_type2} (b). (A) is the JSI before interacting with the sample. (B) is an ideal $\sigma_N=$ 5 nm Notch filter. (D) is a Notch filter with the bandwidth of $\sigma_N=$ 20 nm corresponding to RhB. (F) is a Notch filter with the same bandwidth as RhB, but shifted to the absorption center of this molecule. (C), (E), and (G) show the JSI after passing through filters (B), (D), and (F), respectively.}
	\label{fig:jsi_notch_realistic}
\end{figure*}

The absorption bandwidth of a molecule is generally much broader than 1 nm, and for the case of RhB, it exceeds $20$ nm. In this case, the Notch filter will be so wide that it will completely obscure the JSI distribution provided $\eta$ is high, having no effect on its symmetry and, therefore, being undetectable through the HOM visibility. This phenomenon can be shown in Figure \ref{fig:jsi_notch_realistic}, where different notch filter widths were simulated and their effect on the HOM visibility, $V_\text{out}$, determined. The premise is that maximum decrease of $V_\text{out}$ is correlated with greater asymmetrization of the JSI. The optimal Notch filter bandwidth would be approximately $\sigma_N \approx 5\text{nm}$ provided that 2$\lambda_{p0}$=$\lambda_{N0}$. We can see that for bandwidths  larger than $\sigma_N \geq 10 nm$, the asymmetrizing effect is lost, which is quite unfortunate for the technique we are trying to implement for realistic samples. Nonetheless, detuning $\lambda_{p0}$ from $\lambda_{N0}$ can also induce loss of symmetry. 

In previous sections we had assumed arbitrarily that the central absorption wavelength ($\lambda_{N0}$) matches the double value of the central pump wavelength (2$\lambda_{p0}$). Now we can consider the $\lambda_{N0}$ value for a specific molecule, for instance, RhB, for which $\lambda_{N0} \sim 816$ nm \cite{Triana-Arango2023}. This detuning adds a measurable asymmetry on the JSI (see Figure \ref{fig:jsi_notch_realistic} (F) and (G)). From now on, we will stick with the Notch filter width of $\sigma_N \approx 20$ nm, which matches the actual value for RhB and Rh6G \cite{Makarov2008} for $\Omega_{N_0}=2.31\times10^{15}$ rad/s. 

\subsection{\label{subsec:Theory}Estimation of the Detection Limit}

In the high-intensity regime (classical photons) with pulsed pump, the TPA rate exhibited by a molecule ($R^{(2)}_{C(molec)}=\frac{1}{2}\sigma_C \phi^2$) depends quadratically on the photon flux $\phi$ [cm$^{-2}$ $s^{-1}$], where the classical cross-section $\sigma_C$ is expressed in units of GM, [$1$ GM $= 10^{-50}$ photons cm$^4$ s molecule$^{-1}$]. It is worth noting that for RhB, $\sigma_C$= 120 GM at 800 nm \cite{Makarov2008}. For illustrative purposes, let us calculate the required  excitation to achieve a TPA detection limit of 0.5\% change in the sample transmission, as measured by the standard technique of Z-scan \cite{KuzykCharacterizationtechniques1998}. We assume a short pulse of 100 fs, so the detection threshold requires a pulse energy of 16 nJ, which corresponds to approximately $3.25\times10^{10}$ photons/pulse, assuming a focused beam waist of 30 $\mu$m. To reach such a high number of classical photons, the TPA experiments are typically carried out using bulky generative amplifiers operating at low repetition rate. A low repetition rate with high pulse energy is necessary to avoid thermo-optical effects that can mimic nonlinear electronic responses in Z-scan traces. With a typical repetition rate of 1 KHz, the average incoming photon rate in the sample required to detect TPA is $=3.25\times10^{13}$ photons/s. As will be discussed later, the incoming photon pairs required to reach the detection limit in ETPA is significantly lower and can be achieved using only a high-repetition rate femtosecond oscillator instead of an generative amplifier as in the case of TPA, but appropriate detection approach must be implemented to overcome noise and certificate the ETPA process.    

In the low-intensity pump regime, the ETPA rate for a molecule depends linearly on the photon flux. 
\begin{equation}\label{eq:RateETPA_LinealGrowth}
R_{E (molec)}^{(2)}=\frac{1}{2} \sigma_E \phi,    
\end{equation}

\noindent where $\sigma_E$ is the ETPA cross-section (measured in cm$^2$ molecule$^{-1}$)\cite{Parzuchowski2021}. Now, let us calculate the number of bi-photons absorbed by a sample with a concentration of molecules $C$ [moles/cm$^3$] within an interaction volume $V=A\ell$, with the number of molecules $\#molec=CV N_A$ (where $N_A$ is Avogadro's number). The photon rate is $R^{(1)}_{ph}=N^{(1)}_{ph}/T$ [s$^{-1}$], and the flux is $\phi=N^{(1)}_{ph}/AT=R^{(1)}_{ph}/A$, where $N^{(1)}_{ph}$ is the number of photons passing through a section $A$ in a time $T$. By using these definitions in combination with expression for a single molecule(Equation \ref{eq:RateETPA_LinealGrowth}), we can write the net photon absorption rate as $R^{(2)}=\frac{1}{2} \sigma_E (C N_A \ell)R^{(1)}_{ph}$. If we consider that $R^{(1)}_{in}=2R^{(2)}_{in}$ is the net photon flux from an SPDC process that enters the interaction volume, the expression for the rate of photon pairs absorbed by the sample is finally given by
\begin{equation} \label{eq:RateEtpa_Formulated}
R^{(2)}_{abs}= \sigma_E (C N_A \ell)R^{(2)}_{in}
\end{equation}

The photons exiting the interaction volume (those that were not absorbed) are $R^{(2)}_{out}= R^{(2)}_{in}-R^{(2)}_{abs}$, and the ETPA absorption efficiency, which we introduced earlier, is given by:
\begin{equation} \label{eq:Efficiency_Etpa}
    \eta_E=\frac{R^{(2)}_{abs}}{R^{(2)}_{in}}=\sigma_E C N_A \ell
\end{equation}

This expression will help us determine how the absorption depends on both the intrinsic properties of the molecule given by $\sigma_E$, properties of the photon pair source and the details of the experiment, including the cell thickness, dye under test and sample concentration. In this context, it will be very helpful to determine how many EPPs will be absorbed based on how many pairs are incident on the sample under typical experimental conditions.

The cross-section $\sigma_E$ can be estimated for a given ETPA efficiency under the specific experiment design; by rearranging equation \ref{eq:Efficiency_Etpa}.
\begin{equation} \label{eq:SigmaE_EtaE}
    \sigma_E =\frac{\eta_E}{C N_A \ell}
\end{equation}

For the experimental detection to be statistically significant, the rate of photon pairs lost per second must be greater than the fluctuations $\delta R_{det}$ of the detection system
\begin{equation}\label{eq:detection_limit}
    R^{(2)}_{abs} > \delta R_{det}.
\end{equation}

We can estimate these fluctuations as the standard deviation of the incoming pair rate $R^{(2)}_{in}=\langle R^{(2)}_{in} \rangle+\delta R_{det}$, which for a Poissonian statistic is $\delta R_{det}=\sqrt{R^{(2)}_{in}}$.

When the generated photon pairs are strongly correlated (for example, entangled photons), the fluctuations in the coincidence rate are reduced compared to a Poissonian process due to quantum correlation, $\delta R^2 = F \langle R \rangle$, with $F<1$. We can estimate these fluctuations for a sub-Poissonian statistic as $\delta R_{det}=\sqrt{\frac{1}{2}R^{(2)}_{in}}$ \cite{RARITY1987,IskhakovHeraldedsource2016}.

Thus, the detection limit of our experimental system will depend on the condition in Equation \ref{eq:detection_limit}, where a lower bound is set by the noise fluctuations ($\delta R_{det}$) of the detectors. We can establish a minimum absorption efficiency that the sample must have such that the ETPA can be detected.
\begin{equation}\label{eq:efficiency_min}
    \eta_E^{min}= \frac{\delta R_{det}}{R^{(2)}_{in}}
\end{equation}

Using this minimum ETPA efficiency, $\eta^{(min)}_E$, we can calculate the smallest  cross-section, $\sigma^{min}_E$, which could be measured for a sample, as follows:

\begin{equation}\label{eq:sigmaMINIMA}
    \sigma^{min}_E=\frac{\eta^{(min)}_E}{C N_A \ell}.
\end{equation}

This expression tells us the minimum cross-section that we will be able to detect with our experimental conditions. In order to certify the detection of an experiment, the following conditions, equivalent to Eq (\ref{eq:detection_limit}), must be met
\begin{eqnarray}\label{eq:detection_limit2}
    \eta_E > \eta_E^{(min)} \nonumber \\
    \sigma_E > \sigma_E^{(min)}
\end{eqnarray}

\section{Experimental Methods}

To determine the number of incident photon pairs $R^{(2)}_{in}$ in a standard experimental system intended for ETPA detection, let us consider  the experimental setup depicted in Fig. 6.

\begin{figure*}[t]
	\centering
	\includegraphics[width=0.7\textwidth]{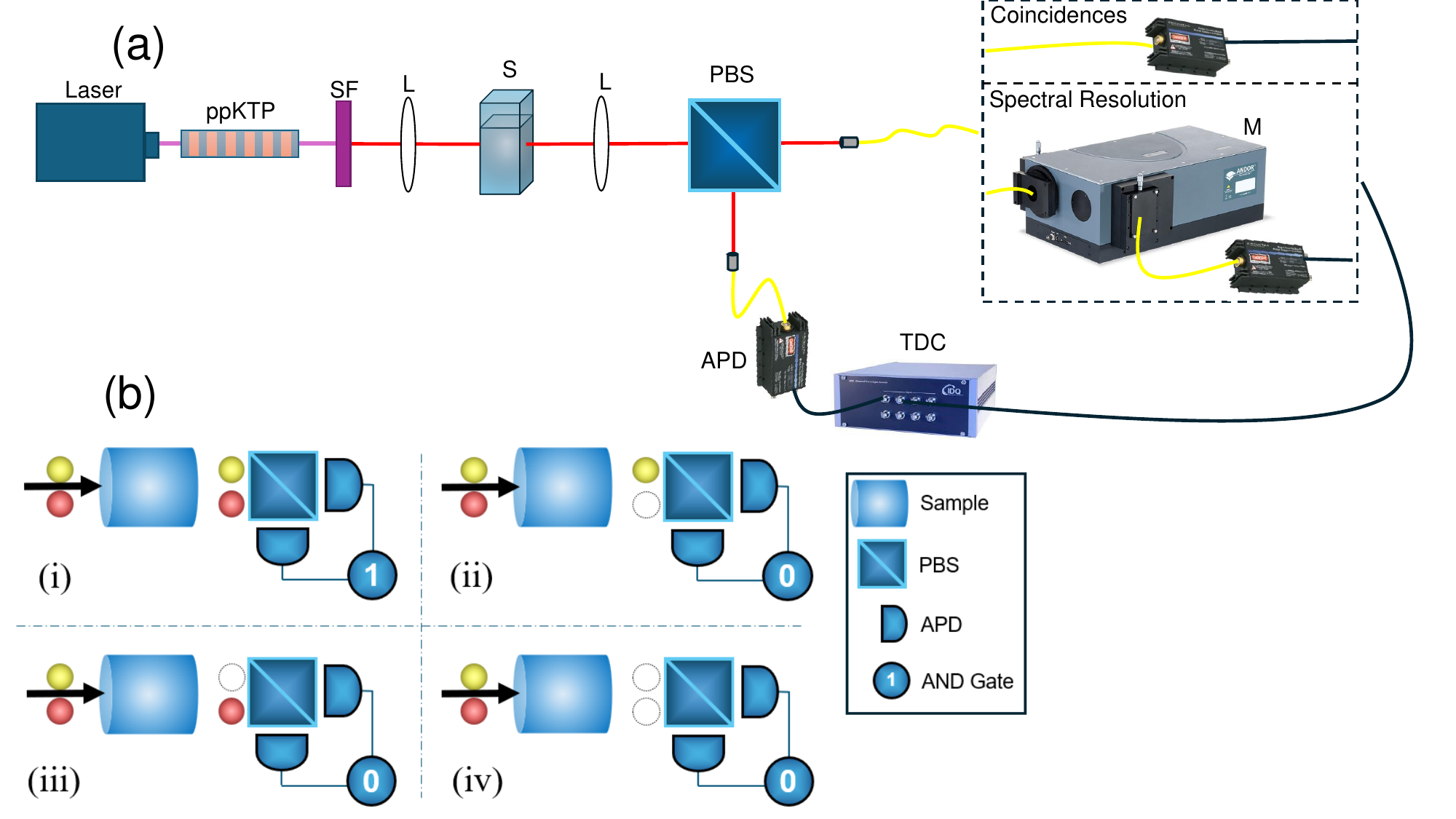}
	\caption{(a) Experimental setup. (b) Possible losses of photons leaving the sample. In this scheme the incident $i$ and $s$ photons are denoted by filled circles, while lost photons after transmission are denoted by empy circles.}
	\label{fig:experimental_array}
\end{figure*}

In Fig. \ref{fig:experimental_array} (a), we show the transmission-based experimental setup used to measure ETPA absorption. We use a Ti:Sa pulsed laser (Chameleon, Coherent Inc.) emitting pulses of $\tau_{puls}=110$ fs, tunable in the wavelength range 650 - 1050 nm, repetition rate of $f_{Rep}$=$80$ MHz, and average power of up to 3 W. The pulsed beam is tuned at 810 nm and doubled in frequency using a BiBO nonlinear crystal, generating photons at 405 nm with a bandwidth of 5 nm and maximum possible power of $P_{avg}=30$ mW. These photons are focused by a lens $L_1$ ($f_1=10$ cm) onto a ppKTP nonlinear crystal (L=10 mm) with a period of $\Lambda=10$ $\mu$m, which generates a type-II SPDC process, with an estimated conversion efficiency of $\epsilon_{spdc}=1.3\times 10^{-9}$. We use spectral filters (SF) to block the pump and allow a bandwidth of 90 nm for the SPDC photons. The SPDC photon flux is focused, via a lens $L_2$ ($f_1=6$ cm), onto a quartz cell (1 cm) containing an RhB sample (S) with concentration $C$, where the transmitted photons are collected by a lens $L_3$ with the same focal length. After the photons exit the sample, they are split by a polarized beam splitter (PBS), where the vertical photon is reflected towards an avalanche photodiode, $APD_1$, and the horizontal photon is transmitted towards two possible options: (i) a second $APD_2$ for coincidence measurements, or (ii) a monochromator and an $APD_3$ which allows to rasterize the frequency of one of the photons in order to obtain the spectral marginal of the JSI. Coincidences are measured in a time-to-digital converter (TDC).

We define the individual coincidence of each detector $R_{1}$, $R_{2}$ and the coincidences $R_{12}$:
\begin{eqnarray}
    R_1 &=& \beta_1 R^{(2)}_{in}+\delta_1 \nonumber \\
    R_2 &=& \beta_2 R^{(2)}_{in}+\delta_2 \nonumber \\
    R_{12} &=& \beta_1 \beta_2 R^{(2)}_{in} +\delta_{12}
\end{eqnarray}

\noindent where $\beta_1$, $\beta_2$ account for all losses, including coupling and detector efficiency, while $\delta_1$, $\delta_2$ are the noise or dark counts for each of the detectors, and $\delta_{12}$ is the accidental coincidence rate. By correcting for the noise counts, we can write $\hat{R}_{\nu} = R_{\nu} - \delta_{\nu}$ ($\nu=\{1,2,12\}$) and solve for the incident photon rate.
\begin{eqnarray}
R^{(2)}=\frac{\hat{R}_{12}}{\hat{R}_{1}\hat{R}_{2}}   \end{eqnarray}

We were able to obtain experimentally the following parameters: $\beta_{i} = 0.21$, $\delta_{i} = 200$ C/s, and the detected single-count rates are $\hat{R}_i = 1.67 \times 10^{7}$ C/s, while the detected coincidence rate is $\hat{R}_{12} = 3.52 \times 10^{6}$ CC/s, with $\delta_{12} = 10$ CC/s.

The advantages of using two-photon correlation detection include robustness to fluctuations and the reduction of uncorrelated sources. In Figure \ref{fig:experimental_array}(b), we show the coincidence logic. Of the four possibilities, only (i) involves both EPPs passing directly through the sample, being separated by polarization, and then detected by two APDs in coincidence. Cases (ii) and (iii) can be associated with linear absorption or scattering losses in one of the detectors while (iv) could represent simultaneous linear losses in both detectors or the elusive ETPA.

\subsubsection{Quantitative Estimation of Photon Absorption}
The energy per pulse $E_{pulse}=P_{avg}/f_{Rep}=3.75\times10^{-10} J/pulse$. The energy of a photon with a wavelength of $405$ nm is $E_{photon}=hc/\lambda$ $=4.9\times10^{-19}$ J. We calculate the peak photon rate used for the excitation of the ppKTP nonlinear crystal, as $R_{peak}=P_{peak}/E_{photon}$ $=6.95\times10^{21}$ photons/s.

Now, to calculate the rate of SPDC photon pairs generated, we multiply the peak photon rate of the pulse at 405 nm by the generation efficiency $R^{(2)(peak)}_{spdc}=R_{peak}\epsilon_{spdc}$ $=9.08\times10^{12}$ pairs/s. To calculate the number of photons per pulse $R^{(2)(perPulse)}_{spdc}=R^{(2)(peak)}_{spdc}\tau_{puls}=0.999$ pairs/pulse. This means that each pulse, on average, produces one pair. Finally, to get the average number of photons per second, we multiply by the repetition rate $R^{(2)(av)}_{spdc}=R^{(2)(perPulse)}_{spdc} f_{Rep}$ $=7.99\times10^{7}$ pairs/s. This value would be equal to the incoming photon rate in the sample $R^{(2)}_{in}=R^{(2)(av)}_{spdc}$. 

\section{Discussion and Results}

Based in our analysis, we can calculate the ETPA rate and the spectral output that would result considering attainable values of EPP fluxes, dark counts of detectors, and real spectral characteristics for a sample. As we discussed in previous sections, we aim to identify at the JSI and the marginals $f(\omega_s)$ and $ f(\omega_s)$ the distinctive signature of the ETPA process, i.e., asymmetric spectral distributions upon absorption process. Here, we proceed taking into consideration  the upper bounds of ETPA cross sections that some authors have estimated recently for representative organic compounds. Table \ref{table: cross sections}  compiles  some of these upper bounds. Let us calculate first the ETPA rate for the case of Rh6G for which an upper bound cross-section of $\sigma_E=1.2\times10^{-25}$ cm$^2$ molecule$^{-1}$  has been estimated by Parzuchowski $et$. $al$ \cite{Parzuchowski2021}. In these computations the concentration is set to $C=58$ mM with an interaction length of $\ell=$ 1 cm. Given this upper bound, the absorption efficiency (Equation \ref{eq:Efficiency_Etpa}) is: $\eta_E=\sigma_E C N_A \ell=4.17\times10^{-6}$. This efficiency implies an ETPA absorption rate (Equation \ref{eq:RateEtpa_Formulated}) $R^{(2)}_{abs}= \eta_E R^{(2)}_{in}=333.81$ pairs/s provided the average number of entangled photon produced by the experimental configuration, depicted in Fig \ref{fig:experimental_array}, is $7.99\times10^{7}$ pairs/s. \\

\begin{table*}[ht]
\begin{center}
\begin{tabular}{ccccccc}
\hline
\hline
                    & C    & $\sigma_C$ & $\sigma_E$                  & $R^{(2)}_{abs}$        & $\eta_E$          \\ 
Sample              & (mM) & (GM)       & (cm$^2$ molecule$^{-1}$)    &  (pairs/s)       &                   \\ 
\hline
$Rh6G$\cite{He2024}       &  $1$      & $---$           & $8\times 10^{-24}$  & $2.2\times10^{4}$ & $2.78\times10^{-4}$& \cite{He2024}  \\
$ICG$\cite{He2024}       &  $1$      & $---$           & $6\times 10^{-23}$  & $1.66\times10^{5}$ & $2.1\times10^{-3}$& \cite{He2024}  \\
\hline
$Rh6G$\cite{Parzuchowski2021}      &$1.5$      & $51$           & $1.2\times 10^{-25}$  & $333.81$ & $4.17\times10^{-6}$& \cite{Parzuchowski2021} \\
$AF455$\cite{Parzuchowski2021}     &$1.5$      & $660$           & $2.1\times 10^{-25}$  & $584.17$ & $7.30\times10^{-6}$ & \cite{Parzuchowski2021} \\
$Qdot$\cite{Parzuchowski2021}      &$0.008$    & $46,000$     & $480\times 10^{-25}$  & $1.33\times10^{5}$ & $1.70\times10^{-3}$& \cite{Parzuchowski2021} \\
$Fluorescein$\cite{Parzuchowski2021} &$1.1$   & $13$           & $1.0\times 10^{-25}$  & $278.17$ & $3.48\times10^{-6}$& \cite{Parzuchowski2021} \\
$9R-S$\cite{Parzuchowski2021}   &$0.39$      & $22$           & $20\times 10^{-25}$  & $5.56\times10^{3}$ & $6.96\times10^{-5}$& \cite{Parzuchowski2021} \\
$C153$\cite{Parzuchowski2021}   &$1.1$      & $14$           & $1.6\times 10^{-25}$  & $445.08$ & $5.56\times10^{-6}$& \cite{Parzuchowski2021} \\
\end{tabular}
\caption{Classical TPA ($\sigma_C$) and estimated upper bounds of ETPA ($\sigma_E$) values for representative molecules. In this table the calculated values of $R^{(2)}_{abs}$ and $\eta_{E}$ result from considering an average rate of incoming photons $R^{(2)}{spdc}=7.9\times10^{7}$ pairs/s. The noise fluctuations for this system is $\delta R_{det}=6,322$ pairs/s. }
\label{table: cross sections}
\end{center}
\end{table*}

In view of the inherent noise fluctuations of the system ($\delta_{det}=6,322$ C/s), which is 19 times greater than $R^{(2)}_{abs}$, it is evident that the ETPA signal is entery obscured. According to Equation \ref{eq:efficiency_min}, the minimum efficiency required to detect ETPA is $\eta_{min}=7.91\times10^{-5}$. Consequently, the smallest cross-section that the system would be able to measure is  $\sigma_E^{min}=2.27\times10^{-24}$ cm$^2$ molecule$^{-1}$, as given by Equation \ref{eq:sigmaMINIMA}. Notice that He $et$. $al.$\cite{He2024} suggested an upper bound for the ETPA cross sections of $\sigma_E=8\times10^{-24}$ cm$^2$ molecule$^{-1}$ (for Rh6G) that surpasses the given $\sigma_E^{min}$ value of the system. In such a case, the limit of detection of the system would be sufficient to detect the ETPA process. 

\subsubsection{Spectral transmitted-based Simulation and Experiment}

\begin{figure*}[t]
	\centering
	\includegraphics[width=0.8\textwidth]{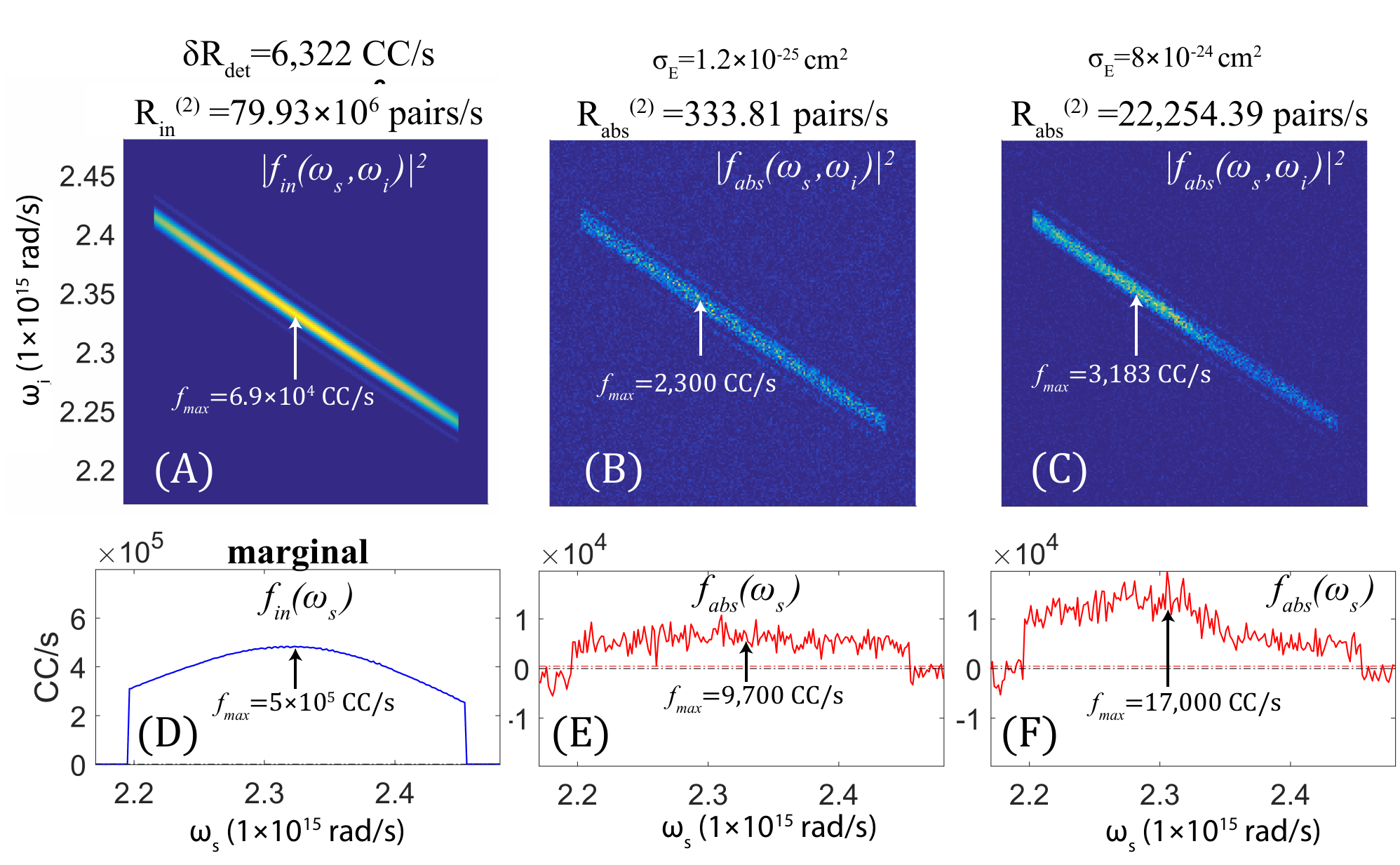}
	\caption{Simulation of the effect of Rh6G using two different cross-sections from the literature in Table 1.
(A) JSI before interacting with the sample.
(B) and (C) Absorbed JSI $\abs{f_{abs}(\omega_s,\omega_i)}^2=\abs{f_{in}(\omega_s,\omega_i)}^2-\abs{f_{out}(\omega_s,\omega_i)}^2$ for ETPA cross sections estimated as upper bounds by Parzuchowski \cite{Parzuchowski2021} and He \cite{He2024}, respectively. (D), (E) and (F) are the marginals of (A), (B) and (C), respectyively.}
	\label{fig:simulation_Rd6G}
\end{figure*}

Now the spectral outputs can be simulated for the cases when the ETPA absorption rate is below or above the limit of detection. Figure \ref{fig:simulation_Rd6G} display the results. Again, we consider the case of Rh6G with the upper bounds enunciated in Table \ref{table: cross sections}, with photon input of $R^{(2)}_{in}=7.99\times10^6$ pairs s$^{-1}$ impiging in 1 cm long cuvette, where the beam is focused to approximately $30$ $\mu$m, achieving an SPDC peak photon flux $\phi$= $1.28\times10^{18}$ pairs cm$^{-2}$ s$^{-1}$. The first column of Figure \ref{fig:simulation_Rd6G} shows the JSI at the input (A) and its corresponding horizontal marginal $f_{in}(\omega_s)$ (D) for a central wavelength of $\lambda_{p0}=405$ nm ($\omega_{p0}=4.65 \times10^{15}$ rad/s), with bandwidth $\sigma_p=5$ nm. A Gaussian noise background was used to predict the fluctuations of the detectors while the final signal is the accumulation of 100 measurements. 

The second column (B and E) in Figure \ref{fig:simulation_Rd6G} refers to the part of the JSI that was absorbed by Rh6G considering the upper bound of cross-section  $\sigma_E=1.2\times10^{-25}$ cm$^2$ molecule$^{-1}$ estimated by Parzuchowski $et$. $al.$ \cite{Parzuchowski2021}, calculated as $\abs{f_{abs} (\omega_s,\omega_i)}^2=\abs{f_{in}(\omega_s,\omega_i)}^2-\abs{f_{out}(\omega_s,\omega_i)}^2$, with $\abs{f_{out}}^2$ being the transmitted JSI. Since $R^{(2)}_{abs}= 333.81$ pairs/s does not exceed the net fluctuation level of $\delta R_{det}=6,322$ pairs s$^{-1}$, the ETPA absorption signature could not be distinguished, even when accumulation was performed on an ensemble of 100 theoretical measurements. We can see from Figure \ref{fig:simulation_Rd6G}(E) that the marginal upon the absorption process falls within the range of noise fluctuations, and as we used a spectral filter, we are left with the shape of the rect filter; this could also be identical if the system had presented linear losses. In contrast, the third column (C and F) of Figure \ref{fig:simulation_Rd6G} presents the computing for Rh6G with the more optimistic threshold of cross-section $\sigma_E=8\times10^{-24}$ cm$^2$ molecule$^{-1}$ estimated by He $et$. $al$ \cite{He2024}. In this case, the absorption rate was improved to $R^{(2)}_{abs}= 22, 254$ pairs/s pairs s$^{-1}$, achieving a signal-to-noise ratio SNR=$3.52$ (12.58 dB) after 100 accumulations. The asymmetric accumulation in the marginal is what gives us the trace that certifies a ETPA spectrum, as demonstrates Fig. \ref{fig:simulation_Rd6G}(F). The fifth and sixth columns in Table 1 summarize the expected values of $R^{(2)}_{abs}$ and $\eta_{E}$ considering the upper bounds of $\sigma_E$ estimated by He $et$. $al.$ \cite{He2024} and Parzuchowski $et$. $al.$ \cite{Parzuchowski2021} for Rh6G, including also the estimated upper bounds and ETPA efficiencies for other representative molecules. In all those cases, the ETPA certification could be performed spectrally when $R^{(2)}_{abs}$ > $\delta R_{det}$.

\begin{figure}[t]
	\centering	\includegraphics[width=0.5\textwidth]{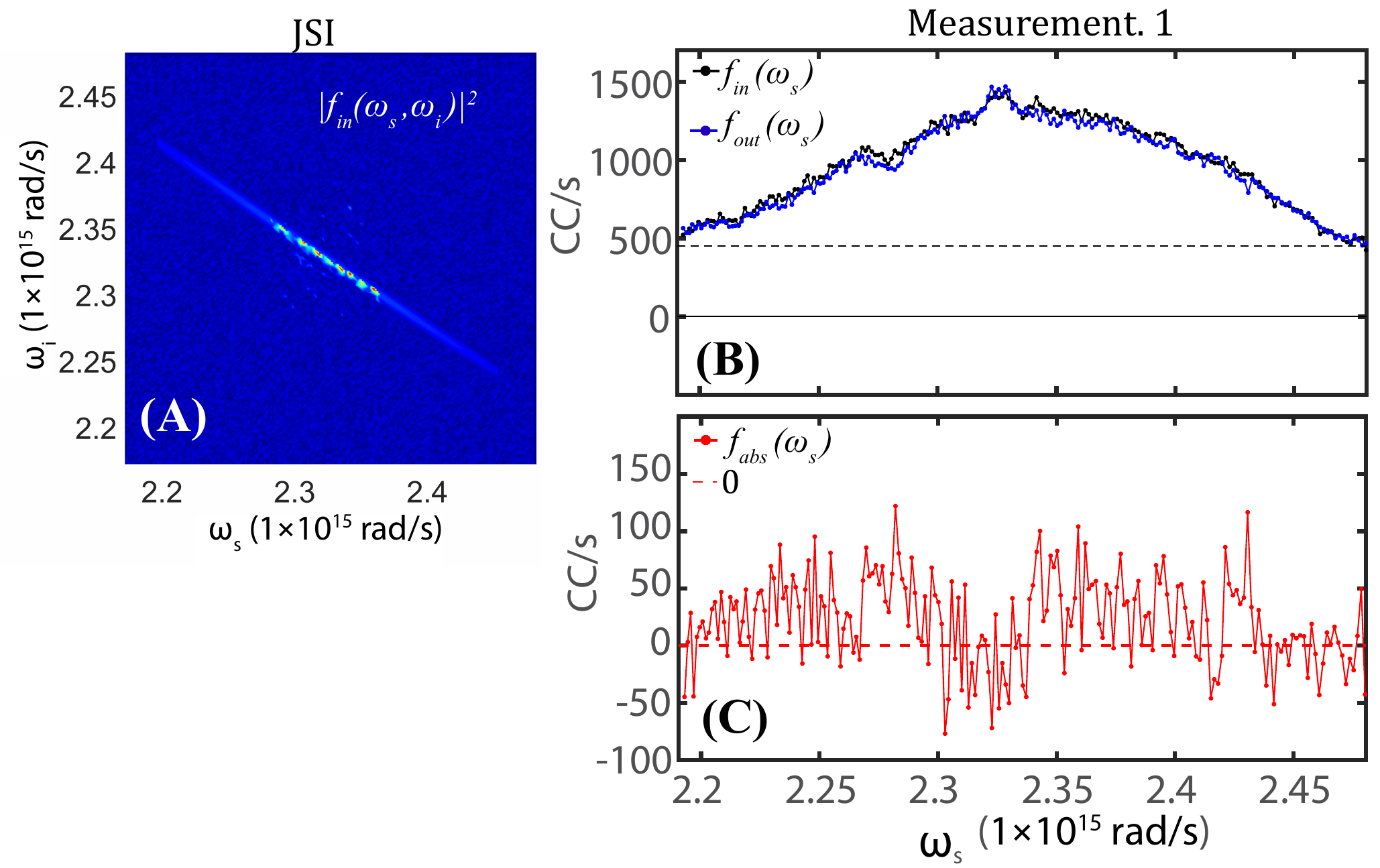}
	\caption{(A) Experimental measurement of the JSI fitted with our theory. (B) [black/blue] Marginal before/after the sample using Methanol as solvent. (C) The absorbed marginal.}
	\label{fig:experimental results}
\end{figure}

Finally, preliminary experimental measurements were made based on the setup shown in Figure \ref{fig:experimental_array}(a). 
Let us recall that the minimum cross section we will be able to measure is determined by the detector fluctuations and by the EPPs rate of the experimental setup (Equations \ref{eq:efficiency_min} and \ref{eq:sigmaMINIMA}), which in our case was $\sigma_E^{min}=2.27\times10^{-24}$ cm$^2$ molecule$^{-1}$. Figure \ref{fig:experimental results} presents the results for RhB in  Methanol with $C=10$ mM. The JSI was taken by adding an extra monochromator to each arm, performed in a localized area, and extrapolated using our theoretical simulations, showing perfect agreement in the slope (panel A). In the second column, the incident and transmitted marginals, and the absorbed marginal are presented (panels B and C, respectively). This measurement may indicate losses due to linear absorption, as it inherits the shape of the incident marginal and no asymmetry is seen other than the shape of the incident spectrum. We can conclude that the ETPA effect was not observed because the cross section of RhB dissolved in methanol must be much smaller than our experimental detection limit of $\sigma_E^{min}=2.27\times10^{-24}$ cm$^2$ molecule$^{-1}$

\section{\label{sec:Conclusion}Conclusions}

A technique based on the spectral resolution of the bi-photon is presented to certify the presence of ETPA through the effect of asymmetric absorption. It is shown that this opens the possibility of performing Hong-Ou-Mandel interferometry once the conditions of symmetry and sensitivity are met. The first condition is that the absorption efficiency must be greater than the threshold efficiency, taking into account the detection limit of the measuring devices. The second condition is that the molecule's absorption spectrum must break the symmetry of the JSI. We expect this technique to be implemented as a reference for designing optimized experiments for the efficient detection of this process.

\section{\label{sec:Associated Content}Associated Content}
\subsection{\label{sec:Data Availability Statement}Data Availability Statement}
        Data underlying the results presented in this paper are not publicly available at this time but may be obtained from the authors upon reasonable request.

\section{\label{sec:Author Information}Author Information}
    \subsection{\label{sec:Corresponding Authors}Corresponding Authors}
    \begin{itemize}
        \item \textbf{Pablo Yepiz-Graciano} $-$ \textit{Instituto de Ciencias Nucleares, Universidad Nacional Aut\'{o}noma de M\'{e}xico, Apdo. Postal 70-543, Ciudad de M\'{e}xico 04510, Mexico}; Email: pablodyg@gmail.com
        \item \textbf{Roberto Ramírez-Alarcón} $-$ \textit{Centro de Investigaciones en Óptica A.C., A. P. 1-948, 37000 León, Gto, Mexico}; Email: roberto.ramirez@cio.mx
        \item \textbf{Gabriel Ramos-Ortiz} $-$ \textit{Centro de Investigaciones en Óptica A.C., A. P. 1-948, 37000 León, Gto, Mexico}; Email: garamoso@cio.mx
    \end{itemize}

\section{\label{sec:Disclosures}Disclosures}
The authors declare no conflicts of interest.

\section{Acknowledgments}
We acknowledge support from SECIHTI, México. This work was supported by SECIHTI, México (CF-I-2699, BP-BSNAC-20250429132744996-10682132).

\bibliographystyle{apsrev4-2}
\bibliography{references}

\end{document}